\documentclass[%
 reprint,
nofootinbib,
nobibnotes,
amsmath,amssymb,
aps,
prd,
]{revtex4-2}

\usepackage{graphicx}%
\usepackage{dcolumn}%
\usepackage{bm}%
\usepackage[separate-uncertainty = true,multi-part-units=single]{siunitx}
\usepackage{hyperref}
\hypersetup{
    colorlinks=true,
    linkcolor=blue,
    citecolor=blue, 
    filecolor=magenta,      
    urlcolor=blue,
    pdftitle={Axion Photon conversion Jesse Satherley},
    }
\usepackage{realboxes}
\usepackage{enumitem}
\usepackage{placeins}
\usepackage[caption=false]{subfig}
\usepackage{tabularx}
\usepackage{multirow}

\usepackage{tikz}
\usepackage{tikz-3dplot}
\usepackage{Functions_Sphere}
\usetikzlibrary{arrows, decorations.markings}
\usetikzlibrary{arrows.meta}
\tikzset{viewport/.style 2 args={
    x={({cos(-#1)*\RadiusSphere cm},{sin(-#1)*sin(#2)*\RadiusSphere cm})},
    y={({-sin(-#1)*\RadiusSphere cm},{cos(-#1)*sin(#2)*\RadiusSphere cm})},
    z={(0,{cos(#2)*\RadiusSphere cm})}
}}

\def\Rotation{160}
\def\Tilt{15}
\def\RadiusSphere{2}
\def\VectorLen{1.5}

\newcommand{\vect}[1]{\bm{#1}}

\newcommand{\unitvect}[1]{\,\mathbf{\hat{#1}}}

\newcommand{\microeV}{\,\mu\text{eV}}
\newcommand{\figureRef}[1]{Fig.~{#1}}

\newcommand{\citelink}[2]{\hyperlink{cite.#1}{#2}} %

\newcommand{\orcidicon}{\includegraphics[width=0.32cm]{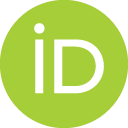}}
\newcommand{\orcid}[1]{\href{https://orcid.org/#1}{\orcidicon}}

\begin{document}

\preprint{APS/123-QED}
\title{Axion-Photon Conversion Signals from Neutron Stars with Spacetime Curvature Accounted for in the Magnetosphere Model}

\author{Jesse C.~Satherley\orcid{0009-0009-2517-2080}}
 \email{jesse.satherley@pg.canterbury.ac.nz}
\author{Chris Gordon\orcid{0000-0003-4864-5150}}%
 \email{chris.gordon@canterbury.ac.nz}
 \affiliation{%
 School of Physical and Chemical Sciences, University of Canterbury, Christchurch, New Zealand
 }
\author{Chris Stevens\orcid{0000-0002-0614-4879}}%
 \email{chris.stevens@canterbury.ac.nz}
 \affiliation{%
 School of Mathematics and Statistics, University of Canterbury, Christchurch, New Zealand
 }%

\date{\today}%

\begin{abstract}
Axions are a well-motivated dark matter candidate. 
They may be detectable from radio line emission due to resonant conversion in neutron star magnetospheres. 
While radio data collection for this signal has begun, further efforts are required to solidify the theoretical predictions for the resulting radio lines. Usually, the flat spacetime Goldreich-Julian model of the neutron star magnetosphere is used, while a Schwarzschild geometry is assumed for the ray tracing. 
We assess the impact of incorporating the spacetime curvature into the magnetosphere model.
We examine a range of neutron star and axion masses and 
find an average difference of $37\%$ and $22\%$ in radiated power compared to the standard Goldreich-Julian magnetosphere model for a $1\microeV$ and $10\microeV$ mass axion, respectively, in the case of a $2.2M_\odot$ mass neutron star. 
A much lesser difference is found for lower-mass neutron stars, as in that case axion-photon conversion occurs further from the Schwarzschild radius. 
\end{abstract}
 
\maketitle

\section{\label{sec:introduction} Introduction}

The axion was introduced to solve the strong CP problem in quantum chromodynamics \cite{pecceiCPConservationPresence1977, pecceiquinn1977, Weinberg1978, Wilczek1978}. 
It was subsequently realised that it could account for a fraction, or all, of the dark matter in the Universe \cite{preskillCosmologyInvisibleAxion1983,abbottCosmologicalBoundInvisible1983,Dine1983}.

A recent and compelling proposal for indirectly searching for axions in astrophysical environments involves detecting radio photons produced by axion-photon mixing in neutron star magnetospheres. The strong magnetic fields and ambient plasma in these regions can resonantly amplify the mixing process, potentially revealing axions through distinctive radio signatures
\cite{pshirkovConversionDarkMatter2009, Huang2018, hookRadioSignalsAxion2018,  Safdi19,leroyRadioSignalAxionphoton2020, fosterGreenBankEffelsberg2020, darlingSearchAxionicDark2020, 
Edwards2021,
battyeRadioLineProperties2021, witteAxionphotonConversionNeutron2021, 
millarAxionphotonConversionStrongly2021,Foster22, noordhuisNovelConstraintsAxions2022, zhouSearchingAxionDark2022, battyeSearchingTimeDependentAxion2023, mcdonaldGeneralizedRayTracing2023,bhura24,walters2024axionsandromedasearchingminicluster}. 
Two main schemes exist to determine the signal received from the axion-photon conversions: (1) the emitter-to-observer scheme samples the conversion surface 
 and propagates photons forward through the plasma \cite{witteAxionphotonConversionNeutron2021}; and
(2) the observer-to-emitter scheme sources photons at a distant detector and propagates them backwards onto the conversion surface \cite{leroyRadioSignalAxionphoton2020}. 
Both methods have their own benefits and drawbacks. 

Early simulations of the axion-photon conversion process assumed a flat spacetime, did not consider the refractive effects of the plasma, and employed a simple flat-spacetime Goldreich-Julian (GJ) model (see Ref.~\cite{goldreichPulsarElectrodynamics1969a}) for the NS magnetosphere (e.g. Refs.~\cite{hookRadioSignalsAxion2018, leroyRadioSignalAxionphoton2020}). 
More recently, work has been done to improve the simulations by including the dispersive effects of an isotropic (unmagnetised) plasma in Schwarzschild spacetime \cite{battyeRadioLineProperties2021}. 
However, this still included the GJ model.
Other studies attempt to account for the random infall of axions onto the neutron star (NS) via a Monte Carlo style simulation \cite{witteAxionphotonConversionNeutron2021}.
Although they include a magnetised plasma dispersion relationship for the photons, they only consider flat spacetime.
Most recently, work was done to include a magnetised plasma dispersion relationship for the photons in Schwarzschild spacetime \cite{battyeSearchingTimeDependentAxion2023}. 
This article also had the inclusion of using a recently derived axion-photon conversion probability incorporating 3D effects \cite{millarAxionphotonConversionStrongly2021}.
However, all these models rely on the inclined GJ model to describe the NS magnetosphere.

The GJ magnetosphere model assumes a dipole magnetic field %
in a flat spacetime. %
In the force-free NS model case, GJ shows that a NS must have a dense magnetosphere containing charged particles.
They derived an analytical expression for the charge density and magnetic field vector around the NS.
The GJ model is expected to be a good description of the closed-field regions of a NS in the flat spacetime approximation \cite{Philippov_2015, Hu_2022}.
Recently, strides have been made to create models that incorporate spacetime curvature. 
Numerical simulations can account for the complete Maxwell equations in the 3+1 formalism of a stationary background metric \cite{petriTheoryPulsarMagnetosphere2016}.
A series of papers by Gralla et al. provide an analytical solution in the near field of an inclined dipole magnetic field in curved spacetime \cite{grallaSpacetimeApproachForcefree2014, grallaElectromagneticJetsStars2016, grallaInclinedPulsarMagnetospheres2017a, lockhartXrayLightcurvesRealistic2019}. 
Their model\footnote{\label{glp_note}Hereafter referred to as the Gralla, Lupsasca, and Philippov (GLP) model.} provides corrections to the dipole magnetic field in GR, which alters the magnetic field strength and shape. 
Their article also includes a charge distribution constituting a plasma in the magnetosphere that depends on GR effects.

While our principal methods follow the work carried out in Refs.~\cite{battyeRadioLineProperties2021, witteAxionphotonConversionNeutron2021, mcdonaldGeneralizedRayTracing2023}, we wish to extend their results by using the GLP model and investigating the potential effects on the predicted power received from axion-photon conversions near a NS. 
We will do this by producing estimated signals in the emitter to observer scheme simulations.
{In the following work we only consider the results of the simpler isotropic plasma and an aligned rotator (similar to several results presented in Ref.~\cite{mcdonaldGeneralizedRayTracing2023}).
Though implementing an anisotropic plasma and a misalignment will effect the results of the simulations, this work aims to check whether a GR derived magnetosphere has an influence in the simplest case. 
Then it would be expected that the aforementioned elements, which depend strongly on the magnetic field structure, could amplify any differences we observe in this foundational model.}

In Sec.~\ref{sec:NS_model}, we provide a review of the GJ model, the GLP model, and a brief aside on plasma.
In Sec.~\ref{sec:axion_photon_conversion}, we discuss our implementation of the conversion of axions to photons in our simulations.
In Sec.~\ref{sec:GR_dispersion_relations}, the dispersion relationships are provided, which are then used in the ray tracing equations reviewed in Sec.~\ref{sec:Ray_tracing}.
In Sec.~\ref{sec:methods}, we explain our implementation of axion-to-photon conversion simulations around neutron stars using the observer-to-emitter scheme.
Lastly, in Sec.~\ref{sec:results_and_discussion}, we present the results of our simulations and the conclusions that can be made. The article's main text considers the case where the magnetic field and rotation axes are aligned. We include details of the misaligned case in the Appendices as a precursor for future work.

Throughout, we denote 4-dimensional abstract (where no particular coordinate system is specified) and coordinate tensor indices with Latin letters starting from $a,b,c,\ldots$ and $i,j,k,\ldots$ respectively, both in the range 0--3 and 3-dimensional coordinate indices with Greek letters $\mu,\nu,\ldots$ in the range 1--3. The 3-vectors with index range 1,2,3 will also be denoted by a boldface typeset where appropriate. We use the metric signature $(-,+,+,+)$, along with the choice of natural units $c=\hbar=\epsilon_0=1$, unless other units are specified.

\section{\label{sec:NS_model} Neutron Star Models}

As already highlighted in the introduction, the typical choice for a NS magnetosphere model used in axion-photon ray tracing simulations is the GJ model, which assumes an inclined dipole magnetic field in a flatspace time. 
However, we wish to explore the effects on these simulations by including a NS model derived in curved spacetime. 
This section reviews and discusses the two models, with a focus on their implementation into the numerical simulations.

\subsection{\label{sec:GJ_NS_model}Goldreich-Julian Neutron Star}

The Goldreich-Julian (GJ) (see Ref.~\cite{goldreichPulsarElectrodynamics1969a}) assumes that the NS is surrounded by a dense plasma of charged particles in the star's magnetosphere. 
GJ derives forms for the magnetic and electric fields for an aligned rotator. 
The GJ model has since been extended to account for the misalignment angle (see for example Ref.~\cite{rezzollaGeneralRelativisticElectromagnetic2001}). 

The number density of electrons and positrons in the magnetosphere of the GJ model is given by,
\begin{equation}
\label{eqn:gj_number_density}
    n_{\text{GJ}}(\vect{r})=\frac{2\vect{\Omega}\cdot\vect{B}}{|q_e|}\frac{1}{1-\Omega^2r^2\sin^2\theta},
\end{equation}
where $\vect{\Omega}=(2\pi/P_{\mathrm{ns}})\unitvect{z}$ is the NS angular velocity vector with $P_{\mathrm{ns}}$ being the period of rotation and $\unitvect{z}$ the unit vector in line with the rotation axis of the star, $q_e$ the charge of an electron, $r$  the distance from the centre of the NS, and $\theta$ the polar angle from the rotation axis. 
The magnetic field is $\vect{B}$, which will be defined shortly for the GJ magnetic field \eqref{eqn:dipole_mag_field}. 
The regions that return a positive number density are dominated by positrons, whereas regions with a negative number density are dominated by electrons. 

The GJ model described above assumes that the star's interior magnetic field takes the form of a dipole. From this assumption, expressions for the fields can be derived. 
In this case only the near-zone fields are necessary.
The near-zone fields are found by taking leading order terms in the limit as $r\to0$, as long as $r>R_{\text{NS}}$ so that the fields are external of the star's surface, where $R_{\text{NS}}$ is the radius of the NS surface.

\begin{figure}
    \centering
    
    \begin{tikzpicture}
        \begin{scope}[viewport={\Rotation}{15}, very thin]
        
            \draw[dashed,->] (-\VectorLen,0,0) -- (\VectorLen,0,0) node[anchor=north]{$x$};
            \draw[dashed,->] (0,0,0) -- (0,0,\VectorLen) node[anchor=east]{$z$};
            \draw[dashed] (0,0,-\VectorLen) -- (0,0,-1);
            \node at (0, 0, -1.65) {(b)};
            
            \draw[thick,->] (0,0,1) -- (0,0,\VectorLen) node[anchor=west]{$\vect{\Omega}$};
            
            \draw[domain=0-\Rotation:360-\Rotation, variable=\azimuth, smooth] plot (\ToXYZ{90}{\azimuth});
            \draw[domain=0:360, variable=\elevation, smooth] plot (\ToXYZ{\elevation}{\Rotation});
            
            \def\Epolar{45}
            \def\Eazimuth{110}
            
            \draw[domain=0:1, variable=\radius, dashed] plot (\ToXYZr{\radius}{\Epolar}{\Eazimuth});
            \draw[domain=1:\VectorLen, variable=\radius, thick, ->] plot (\ToXYZr{\radius}{\Epolar}{\Eazimuth}) node[anchor=south]{$\vect{e}$};
            
            \draw[domain=0:\Epolar, variable=\elevation, smooth] plot (\ToXYZr{0.4}{\elevation}{\Eazimuth}) node[] at (\ToXYZr{0.5}{22.5}{135}) {$\chi$};
            \draw[domain=0:\Eazimuth, variable=\azimuth, smooth] plot (\ToXYZr{0.4}{90}{\azimuth}) node[] at (\ToXYZr{0.65}{90}{\Eazimuth/2+10}) {$\Omega t$};
            \draw[domain=0:90, variable=\elevation, dashed] plot (\ToXYZr{1}{\elevation}{\Eazimuth});
            \draw[domain=0:1, variable=\radius, dashed] plot (\ToXYZr{\radius}{90}{\Eazimuth});
            
            \draw[domain=\Eazimuth:\Eazimuth+40, variable=\azimuth, smooth] plot (\ToXYZr{0.6}{90}{\azimuth}) node[] at (\ToXYZr{0.8}{90}{\Eazimuth+40/2}) {$\lambda$};
            \draw[domain=0:1, variable=\radius, dashed] plot (\ToXYZr{\radius}{90}{\Eazimuth+40});
        \end{scope}
    \end{tikzpicture}
    \caption{The star's rotation axis $\vect{\Omega}$ and the magnetic field's symmetry axis $\vect{e}$. 
    The angle $\chi$ is the inclination between $\vect{\Omega}$ and $\vect{e}$, and $\lambda = \phi - \Omega t$ is measured from the projection of $\vect{e}$ onto the xy-plane. The magnetic axis $\vect{e}$ has azimuthal angle $\Omega t$ in $(r,\theta,\phi)$ coordinates}
    \label{fig:NS_coordinates}
\end{figure}
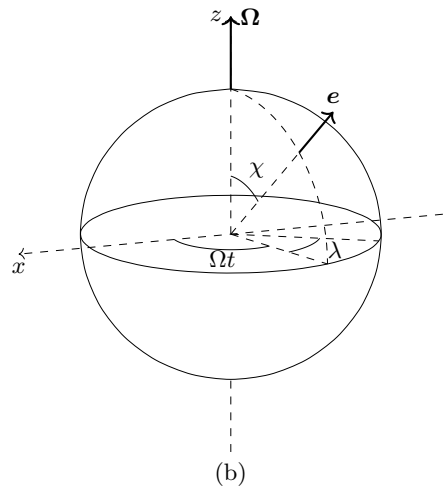

In the GJ model within the near-zone, the magnetic fields are given by that of an idealised inclined rotator (e.g. \cite{rezzollaGeneralRelativisticElectromagnetic2001, satherleyPedagogicalReviewVacuum2022}),
\begin{equation}
\label{eqn:dipole_mag_field}
    \begin{aligned}
        B_r &= \frac{2\mu}{r^3}\left(\cos\chi\cos\theta + \sin\chi\sin\theta\cos\lambda\right),\\
        B_\theta &= \frac{\mu}{r^3}\left(\cos\chi \sin\theta - \sin\chi\cos\theta\cos\lambda\right),\\
        B_\phi &= \frac{\mu}{r^3}\sin\chi\sin\lambda,
    \end{aligned}
\end{equation}
with $\mu$ being the magnetic dipole moment of the star, $\chi$ the misalignment angle between the rotation axis and the magnetic field axis, and $\lambda=\phi-\Omega t$, with $\phi$ the azimuthal angle around the NS where $\phi=0$ is inline with $\unitvect{x}$ and increases in the anticlockwise direction.
These coordinates are shown in Fig.~\ref{fig:NS_coordinates}.

Because the GJ model is the simplest NS model which includes a plasma with a well-defined charge density, it is the usual choice for axion-photon conversion papers, for example \cite{witteAxionphotonConversionNeutron2021, battyeRadioLineProperties2021, battyeSearchingTimeDependentAxion2023, leroyRadioSignalAxionphoton2020, hookRadioSignalsAxion2018, noordhuisNovelConstraintsAxions2022}. 
However, we wish to study the effect of including a dipole magnetic field derived in curved spacetime and compare it to the results using the GJ model.
In the following section we detail necessary equations from the GLP model.

\subsection{\label{sec:GLP_NS_model}Gralla, Lupsasca and Philippov Magnetosphere}

In recent literature, derivations have been published that extend the dipole magnetic field around a NS to a curved spacetime.
The GR pulsar model we explore in this work is taken from a series of papers by Gralla et al. \cite{grallaSpacetimeApproachForcefree2014, grallaElectromagneticJetsStars2016, grallaPulsarMagnetospheresFlat2016, grallaInclinedPulsarMagnetospheres2017a, lockhartXrayLightcurvesRealistic2019}.
In their work, they use the Hartle-Thorne metric \eqref{eqn:hartle_thorne_metric} to describe the electromagnetic fields around a pulsar.
The Hartle-Thorne metric is used to describe a slowly rotating star in general relativity.
It is given as (e.g. \cite{grallaPulsarMagnetospheresFlat2016}),
\begin{equation}
    \label{eqn:hartle_thorne_metric}
    g_{ij} = 
    \begin{bmatrix}
        -f^2+\Omega^2r^2\sin^2\theta & 0 & 0 & -\Omega r^2\sin^2\theta \\
        0 & 1/f^2 & 0 & 0 \\
        0 & 0 & r^2 & 0 \\
        -\Omega r^2\sin^2\theta & 0 & 0 & r^2\sin^2\theta \\
    \end{bmatrix},
\end{equation}
where $f = 1 - r_s/r$ is the Schwarzschild function
and uses Schwarzschild coordinates $x^i = \{t,r,\theta,\phi\}$. 
The rotation is contained with the terms including $\Omega$, which is the angular velocity of the star. 
When the rotation is slow enough, frame-dragging terms that contain $\Omega$ become insignificant.
The Hartle-Thorne metric then simplifies to the Schwarzschild metric. 

In our work, we continue using the Schwarzschild metric for both the magnetic field and the ray tracing\footnote{We checked that the Hartle-Thorne metric does not make a significant difference to our results. However, it may matter for a faster-rotating NS.}.
Gralla et al. provide solutions for the near fields in the situation of a force-free axisymmetric field, after which they extend to an inclined pulsar where the misalignment angle is non-zero. 
In the following, we detail their work, and we provide insights on how to apply their results to the problem of axion photon conversion near a NS.
We begin by describing a set of electromagnetic relationships.

We can define the magnetic field for an arbitrary observer and metric by using (e.g.~\cite{grallaSpacetimeApproachForcefree2014})\footnote{There is a missing factor of a $1/2$ for the equivalent of this equation in Eqn.~(29) of Ref.~\cite{mcdonaldGeneralizedRayTracing2023} and Eqn.~(1) of Ref.~\cite{gedalinWavesStronglyMagnetized2001}.},
\begin{equation}
    \label{eqn:covariant_magnetic_field}
    B^d = \frac{1}{2}\epsilon^{abcd}F_{ab}U_c
\end{equation}
where $F_{ab}$ is the electromagnetic tensor, $U_c$ is the 4-velocity of an observer and $\epsilon^{abcd}$ the covariant Levi-Civita tensor to account for the metric, which is related to the Levi-Civita symbol multiplied by the determinant of the metric tensor such that (e.g.~\cite{carrollSpacetimeGeometryIntroduction2004}),
\begin{equation}
    \begin{aligned}
        \epsilon_{abcd} &= \left(\sqrt{|\det{g}|}\right)\varepsilon_{abcd},\\
        \epsilon^{abcd} &= \frac{\text{sign}\left(\det{g}\right)}{\left(\sqrt{|\det{g}|}\right)}\varepsilon^{abcd},
    \end{aligned}
\end{equation}
where $\varepsilon_{abcd}$ is the fourth-rank Levi-Civita symbol.
The last relationship we require is the current density 4-vector, as the charge density is contained within the current density 4-vector. It is related to the 3-space quantities by the relationship,
\begin{equation}
    J^a=(\rho, \vect{J}),
\end{equation}
where $\rho$ is the charge density and $\vect{J}=\rho\vect{u}$ is the current density  with velocity $u^\mu=dx^\mu/d\tau$, where $\tau$ is the proper time. The current density 4-vector is related to the electromagnetic tensor by,
\begin{equation}
    \label{eqn:4-current_em_tensor}
    J^a = \nabla_b F^{ab},
\end{equation}
where $\nabla_b$ is the covariant derivative. 

\subsubsection{Aligned Rotator}
The first paper in the series by Gralla et al.~\cite{grallaPulsarMagnetospheresFlat2016} begins by deriving an analytical method for studying the force-free magnetosphere of a slowly rotating aligned rotator ($\chi = 0$), including the effects of GR. 
They find that a $\sim60\%$ correction to the dipole component of the surface magnetic field is introduced by accounting for GR. 

In deriving the equations for the magnetic field, they assume that the electromagnetic field is force-free ($F_{ab}J^b=0$).
This is the same assumption as the GJ  model.
The electromagnetic tensor can then be given by potentials $\psi_i=\psi_i(r, \theta, \phi - \Omega t)$, such that,
\begin{equation}
    \label{eqn:electromagnetic_tensor_index_form}
    F_{ij} =  \partial_i\psi_1\partial_j\psi_2 - \partial_j\psi_1\partial_i\psi_2.
\end{equation}
As the field is axisymmetric, the time dependence does not alter the field configuration. 
However, we include it for completeness as it is required when considering the inclined magnetic field case. 
In the case of an aligned rotator, the magnetic flux function for a dipole in the near-field region is given as,
\begin{subequations}
    \begin{align}
            \psi_{1,\;\text{near}}(r, \theta) &= \mu R^>_1(r)\sin^2\theta,\label{eqn:mag_flux_function_1}\\
            \psi_2(\phi-\Omega t) &= \phi-\Omega t\label{eqn:mag_flux_function_2},
    \end{align}
\end{subequations}
where $\mu=B_\mathrm{ns}R_{\rm ns}^3/(2\Delta_1)$ is the dipole moment with %
$B_\mathrm{ns}$ being the surface magnetic field strength at the pole,\footnote{This is related to the magnetic moment of \eqref{eqn:dipole_mag_field} via $B_\mathrm{ns}=\mu$.} $R^>_1(r)$ is the first radial harmonic, and $\Delta_\ell=R_{\rm ns}^\ell R^>_\ell(R_{\rm ns})$, all with $R_{\rm ns}$ the 
radius of the star. 
The function $\Delta_\ell$ is dimensionless and depends only on the compactness of the star. 
It provides the GR correction to the dipole moment at the surface.
The first radial harmonic is (e.g.~\cite{beskinMHDFlowsCompact2010}),
\begin{equation}
    R^>_1(r)=-\frac{3}{2r}\left[\frac{3 - 4f + f^2 + 4\log f}{\left(1-f\right)^3}\right],
\end{equation}
recalling that
$f=1-r_s/r$ is the Schwarzschild function. 

For this model to be beneficial, it must have a magnetosphere containing charged particles.
The charge density around the star can be described by investigating the charge-current 4-vector.
Upon applying the covariant derivative to the electromagnetic tensor with the Hartle Thorne metric \eqref{eqn:hartle_thorne_metric}, to leading order derivative terms, the time component of $J^a$ is\footnote{We have added back in factors of $G$ compared to GLP's derivation due to our choice of units.},
\begin{equation}
    \label{eqn:GLP_aligned_J_t}
    J^t = \frac{2\left(\Omega - \Omega_z\right)}{r\left(r - 2 G M\right)}\left[\left(r - 3 G M\right)\partial_r\psi_1 + \cot\theta\partial_\theta\psi_1\right],
\end{equation}
which with,
\begin{equation}
    \label{eqn:GR_charge_density}
    \rho_e=U_aJ^a=J^t\sqrt{1-\frac{r_s}{r}}=J^{\hat{t}},
\end{equation}
gives the charge density around the NS for the aligned rotator case.
This relationship is used to find the charge density for the plasma frequency, which will be discussed shortly.
We can then compare the results of the fields and charge density described here to the GJ model \eqref{eqn:gj_number_density} and \eqref{eqn:dipole_mag_field}  to see the effect these corrections have on the axion-photon conversion signal.

In the simulations presented in this paper, we take the magnetic field and rotation axis to be aligned so that we may compare with Ref.~\cite{mcdonaldGeneralizedRayTracing2023} (henceforth referred to as \citelink{mcdonaldGeneralizedRayTracing2023}{MW23}).
So, for the results presented in this paper, it is only necessary to understand the GLP-aligned rotator reviewed in the main body.
However, our code implements the inclined rotator case of the GLP model.
In Appendix~\ref{appendix:GLP_inclined}, we detail the equations and relationships necessary to include the inclined GLP model.

\subsection{Plasma}

A fundamental parameter that characterises a plasma is its plasma frequency. In the absence of a magnetic field, the plasma frequency is the oscillation frequency for the charge distribution about its equilibrium and is given as (e.g.~\cite{swansonPlasmaWaves2003})
\begin{equation}
\label{eqn:plasma_frequency}
    \omega_p=\sqrt{\sum_i\frac{4\pi\alpha n_i}{m_i}},
\end{equation}
where $m_i$ and $n_i$ is the mass and number density of species $i$, and $\alpha$ is the fine-structure constant.

It is assumed that the plasma consists only of electrons and positrons.\footnote{If ions were considered for the positively charged regions instead, the plasma frequency would decrease due to the ion's greater mass.} 
This type of plasma is used, for example, by Refs.~\cite{hookRadioSignalsAxion2018, leroyRadioSignalAxionphoton2020, battyeDarkMatterAxion2020, witteAxionphotonConversionNeutron2021, mcdonaldGeneralizedRayTracing2023}. 
The number density of the charged particles and the charge density are just  related via,
\begin{equation}
    \label{eqn:number_density}
    n_e = \frac{|\rho_e|}{q_e},
\end{equation}
where $q_e$ is the charge of an electron.

In previous studies three cases are considered for the plasma:
\begin{enumerate}
    \item The plasma consists of no charged particles in the magnetosphere, considering only a vacuum around the star. 
    \item An isotropic plasma, where charged particles are present, but the magnetic field of the NS has no effect on the medium. In this case, the plasma frequency only depends on the distance from the star.
    \item An anisotropic plasma, where the magnetic field induces new effects on the medium. 
\end{enumerate}
The choice of plasma is important for the ray tracing of photons as it alters their dispersion. 
In this work we only consider the simpler isotropic plasma as an initial study to check whether considering a GR magnetic field has any effect.

\section{Axion-Photon Conversion} \label{sec:axion_photon_conversion}

As previously indicated, axions undergo resonant conversion to photons in the presence of a plasma (e.g.~\cite{millarAxionphotonConversionStrongly2021}).
The benefit of this is an enhancement to the photon signal from axion-photon conversions, potentially leading to detectable signals from Earth.
The resonance occurs due to plasma effects generating an effective mass for the photon, allowing the axion and photon dispersion relations to intersect. 
This resonant condition is maximised when the plasma frequency is close to or equal to the mass of the axion,\footnote{In SI units, this condition is $\omega_p=\hbar m_a$.} %
\begin{equation}
\omega_p \approx m_a\, .
\label{eqn:resonance_condition}
\end{equation}
\citelink{mcdonaldGeneralizedRayTracing2023}{MW23} uses this condition for their isotropic plasma cases.
For an anisotropic plasma, they advocate using full kinematic matching of axion to photon conversion, which is given by $k_a = k_\gamma$.

\subsection{Conversion Surface} \label{subsec:conversion_surface}

Using the resonance condition in \eqref{eqn:resonance_condition},
we can define a surface in three-dimensional space surrounding a magnetic field source with a plasma.
This surface is referred to as the \emph{conversion surface}.
The conversion surface is the region that will have the most predominant axion-photon conversion flux.
When considering the GJ dipole magnetic field, where $B\propto1/r^3$, the charge number density \eqref{eqn:gj_number_density} can have its $r$ dependency explicitly shown as,
\begin{equation}
    n(r, \theta, \phi, t) \approx \frac{1}{r^3}n(r=1\,\text{eV}^{-1}, \theta, \phi, t),
\end{equation}
where we have ignored the rightmost fraction in \eqref{eqn:gj_number_density}, as it is near unity for small radii.
The expression above can then be combined with \eqref{eqn:plasma_frequency} and \eqref{eqn:resonance_condition}
to give the radius of the conversion surface for a given $\theta$, $\phi$ and $t$:
\begin{equation}
    \label{eqn:conversion_surface}
    r = \sqrt[3]{\frac{\omega_p^2(r=1\,\text{eV}^{-1}, \theta, \phi, t)}{m_a^2}}.
\end{equation}
For the GLP model, where the dependency on $r$ is more complicated, root-solving methods are required to find the conversion surface. 

\subsection{Probability of Conversion}

As the process of an axion converting to a photon is based on classical field theory and is mediated by the interaction term in the axion's Lagrangian,  %
an associated probability of conversion can be found (e.g.~\cite{millarAxionphotonConversionStrongly2021}).
This probability will affect the total radiated power predicted in simulations.
Hence, the choice of conversion probability method that is used has a significant impact on the results.
For this reason, we chose to use the conversion probability for an isotropic plasma that was given in Eqn.~(69) of \citelink{mcdonaldGeneralizedRayTracing2023}{MW23} so that we may compare results.
The relationship also conveniently incorporates a curved spacetime.
This conversion probability is expressed as,
\begin{equation}
    P_{a\gamma\gamma} = \pi g^2_{a\gamma\gamma}|\boldsymbol{B}|^2\sin^2\tilde{\theta} \frac{E_\gamma}{|k^i\partial_i(\omega_p^2)|},
    \label{eqn:conversion_probability_isotropic}
\end{equation}
where $g_{a\gamma\gamma}$ is the axion-photon coupling constant, $|\boldsymbol{B}|^2=B_\mu B^\mu$, $\tilde{\theta}$ is the angle between the axion's momentum and the magnetic field, and $E_\gamma$ is the energy of the photon at the point of conversion. 

\section{\label{sec:GR_dispersion_relations}Curved Spacetime Dispersion Relations}

The warping of spacetime from the mass of a NS can be extreme near the star, significantly affecting the path that particles and light would follow compared to a flat spacetime.
By accounting for %
these 
influences on geodesics around the star, the results of axion-photon conversion simulations will be changed (see for example Refs.~\cite{whiteBlacklightGeneralrelativisticRaytracing2022, pelleSkylightNewCode2022}). 

For the curvature of spacetime to be accounted for in the dispersion relationships above, the metric must be included in some form.
A simple method of converting the flat spacetime relationships to a curved spacetime is by simply taking the $3+1$ approach where squared parameters can be converted to Einstein sums, as done in Ref.~\cite{rogersFrequencydependentEffectsGravitational2015}. 
We can, however, extend this further by introducing covariant relationships for the photon 4-vector components.
In the following, we discuss how gravitational effects are accounted for in the photon dispersion relation inside different types of plasma. 

Firstly, the refractive index of the medium takes on a covariant form and becomes (e.g. Ref.~\cite{rogersFrequencydependentEffectsGravitational2015}),
\begin{equation}
    \label{eqn:GR_refractive_index}
    n^2= 1 + \frac{k_ak^a}{(k_bU^b)^2},
\end{equation}
where $k^a$ is the photons 4-momentum\footnote{Remember that in natural units, the 4-momentum and 4-wavevector are equivalent. In SI, $p^a=\hbar k^a$.} and $U^a$ is a global unit timelike vector such that $U^aU_a=-1$, recalling that $U^a$ is the 4-velocity of an observer.
When the Minkowski metric is used, \eqref{eqn:GR_refractive_index} reduces to the flat-space case of {$n=\omega$/k} where $\omega$ is the photon frequency and $k$ is the photon wavenumber.

\subsection{Vacuum}
For the simple case when no plasma is present, only gravity will affect the path the photon travels. 
The dispersion relationship in a vacuum is just, 
    \begin{equation}
\label{eqn:vacuum_dispersion_relation_GR}
        D(k)=k_ak^a=0,
    \end{equation}
where the sum over the indices contains the metric, which will account for the curvature of space.
With the Minkowski metric, this simply becomes $k^2-\omega^2$. 
In this case, photons travel along geodesics around the star.

\subsection{Isotropic Plasma}
With the inclusion of a plasma, there is now a scalar function $\omega_p(x^a)$ present.
It can essentially be considered a forcing term in the dispersion relationship, which alters the trajectory of the photons from the vacuum case. 
The function $\omega_p$ is a scalar independent of the metric. 
Hence, it remains the same as the flat-space case, as it is unaffected by the inclusion of GR. 
In an isotropic plasma, the dispersion relationship is,
\begin{equation}
    \label{eqn:unmagnetised_dispersion_relation_GR}
    D(k)=k_ak^a+\omega_p^2=0.
\end{equation}
For the derivation of this relationship see Eqn.~(10) in Ref.~\cite{rogersFrequencydependentEffectsGravitational2015}, which also matches with  Eqn.~(17) of Ref.~\cite{battyeRadioLineProperties2021}.

\subsection{Anisotropic  Plasma} \label{subsubsec:magnetised_cold_plasma}
The covariant form of the anisotropic dispersion relationship requires some covariant plasma expressions presented in Ref.~\cite{gedalinWavesStronglyMagnetized2001}\footnote{They use a metric with the opposite signature $(+, -, -, -)$. We modify their relationships to be compatible with our choice of metric signature, which is $(-, +, +, +)$. This leads to a difference in signs on some terms.}.
As in previous sections, let $U^a$ be a global unit timelike vector such that $U^aU_a=-1$.
Then, the photon's effective energy measured by an observer with 4-velocity $U^a$ is $W=-k_aU^a$.
Also, define the unit vector in the direction of the magnetic field $b^a = B^a/\sqrt{B^cB_c}$ with $b^ab_a=1$ (where in Schwarzschild coordinates $B^0=B^t=0$). 
Then, the wave vector component parallel to the magnetic field can be represented by the sum $K_\|=k_ab^a$. 
In the non-relativistic plasma limit of Eqn.~(12) of Ref.~\cite{gedalinWavesStronglyMagnetized2001}, we have that the GR anisotropic dispersion relation is,
\begin{equation}
\label{eqn:magnetised_dispersion_relation_GR}
    D(k)=k_ak^a+\omega_p^2\left(1 - \frac{K_\|^2}{W^2}\right)=0.
\end{equation}
When the appropriate Minkowski limit is taken, where $K_\|=k\cos\tilde{\theta}$ and $W=\omega$, the above equation simplifies to the flat-space case.

\section{Ray Tracing} \label{sec:Ray_tracing}
When a photon propagates through a plasma, it may undergo refraction and reflection due to the plasma's varying refractive index. To trace the path of the photons through the plasma, \emph{ray tracing} is used. 
At a simple level, this involves a system of coupled \emph{ordinary differential equations} (ODEs), which are constructed using one of the plasma dispersion relationships in Sec.~\ref{sec:GR_dispersion_relations}. 
The ODEs `tell' the photon which direction to travel, how its momentum should change direction, and how its energy should evolve. 
The ODEs can then be solved analytically or integrated using a numerical solver, depending on the complexity of the dispersion relation that forms the ODEs. 
The solutions allow the photon to be followed through the magnetosphere of the NS and can be used to reproduce the expected photon signal from axion-photon conversions.
When the dispersion relation contains a plasma frequency term, the ray path is most affected when the frequency of a photon is close to the plasma frequency.

\subsection{Ray Tracing in General Relativity} \label{subsec:GR_ray_tracing}

Paths of geodesics in flat space-time are simply straight lines. In a curved spacetime additional terms involving the Christoffel symbols appear in the geodesic equation, altering the paths \cite{grassoGeometricOpticsGeneral2019, isaacsonGravitationalRadiationLimit1968}. In a curved spacetime with the presence of a plasma, the paths will change further \cite{gedalinWavesStronglyMagnetized2001}. Below we give the equations governing null geodesics in a curved spacetime with plasma.

\subsubsection{Geometric Optics \label{subsubsec:GEO_GR_ray_tracing}}
The geometric optics limit is formed by taking the Wentzel–Kramers–Brillouin approximation with the eikonal form \cite{gedalinWavesStronglyMagnetized2001}. The general relativistic equations for ray propagation can then be found by first representing a wave packet in the form,
\begin{equation}
    A_b=\int \Bar{A}_b(k)e^{ik_bx^a}\sqrt{|g|}d^4k
\end{equation}
having used the eikonal form with a Fourier transform. In the exponential, $k_b$ should satisfy $D(k,x)=0$, and in particular, there should be a sharp maximum when $k=k_0$. So, we can make the substitution $k=q+k_0$. Then, along the ray $x^a=x^a(\lambda)$, where $\lambda$ is an affine parameter along the light path, the phase $q_ax^a$ should be stationary. So, we will have $q_a(dx^a/d\lambda)=0$. One can then expand the dispersion relation about $k_0$ to obtain $q_a(\partial D/\partial k_a)=0$. This yields,
\begin{equation}
    \frac{dx^a}{d\lambda} = \frac{\partial D}{\partial k_a}.
\end{equation}
Then using $dD/d\lambda=0$,
\begin{equation}
    \frac{\partial D}{\partial k_a}\frac{dk_a}{d\lambda} + \frac{\partial D}{\partial x_a}\frac{dx^a}{d\lambda} = 0.
\end{equation}
Hence, we will have a system of ODEs that describes the ray path. They are expressed as,
\begin{subequations}
    \begin{align}
        \frac{dx^a}{d\lambda} &= \frac{\partial D}{\partial k_a}, \label{eqn:GR_geometric_equations_a} \\
        \frac{dk_a}{d\lambda} &= -\frac{\partial D}{\partial x^a}.\label{eqn:GR_geometric_equations_b}
    \end{align}
\end{subequations}
The solution to the first equation \eqref{eqn:GR_geometric_equations_a} prescribes the spacetime position, whilst the second equation \eqref{eqn:GR_geometric_equations_b} controls the energy and direction of propagation.
The convenience of this form is once we have defined the dispersion relation in covariant form, we may then directly use it to find the ray paths. The only caveat is this method requires the dispersion relation to have the 4-wavevector expressed as a covariant vector due to the derivative present in \eqref{eqn:GR_geometric_equations_a} while the position 4-vector remains in contravariant form.

\section{Methods} \label{sec:methods}

In this section, we describe our procedure to produce the expected axion-photon conversion signals around a NS.
This section begins with the required numerical methods to complete the simulations.
This ultimately allows us to compare the effects of changing the metric, dispersion relation, magnetic field, and conversion probability on the signal received from axion to photon conversion around a NS.

\subsection{Numerical Methods}

The use of numerical methods are employed to traverse the photons through the plasma surrounding the star, find their intercepts with the conversion surface (the axion-photon conversion points), and produce the estimated signal/flux from these phenomena. 
Our choice of programming language is Python\footnote{\url{https://www.python.org/}} due to its relative simplicity and an exhaustive library of modules capable of running the simulations carried out in this work.

Due to the complexity of the coupled ODE systems present from the dispersion relations and the ray tracing equations, we use numerical solvers to compute the paths of the photons through the plasma. 
We employed the Scipy library's solve\_ivp function to numerically trace the position and momentum of the photons through the dispersive plasma.
At each step, the magnetic field, plasma frequency, and angle between the magnetic field and the momenta of the photons is computed.

To compute the derivatives of functions, the Python library Autograd\footnote{\url{https://github.com/HIPS/autograd}} was employed. Autograd is capable of automatically differentiating native Python and Numpy code. 
It works by using \emph{Automatic Differentiation} to compute an approximation of the derivative of a function, with machine precision accuracy, without computing a symbolic expression of the derivative.
Hence, only the function must be known, but its related derivatives are unnecessary.
If the plasma function is to be modified, rather than finding its potentially difficult derivative, only the function itself needs to be known for Autograd to return the derivative/gradient of the plasma function in each coordinate direction.

We implement the module Multiprocessing\footnote{\url{https://docs.python.org/3/library/multiprocessing.html}} to leverage the multiple threads available to us during a simulation by propagating multiple photons through the plasma simultaneously. 
This allowed us to divide the time for a simulation to run by approximately the number of available processing threads. 

The code was developed so that changes to the dispersion relation, magnetic field, or probability calculations can be easily made by modifying the defining Python functions. 
This also allows for adding more complex magnetic fields, dispersion relations, or conversion probability relationships.

\subsection{Observer-to-Emitter Scheme}

{In the observer-to-emitter scheme, the photons are first sourced at the image plane, then they are propagated backwards along their trajectory \cite{luminetImageSphericalBlack1979, pelleSkylightNewCode2022, schnittmanHarmonicStructureHigh2004, nobleSimulatingEmissionOutflows2007, dexterFastNewPublic2009}. 
This concept takes the physical process of photons being generated at a source, after which
the photons propagate towards and become incident on a detector placed asymptotically far from the star.
The difference being the observer-to-emitter scheme reverses the trajectory.
Due to the assumption that the source is asymptotically far from the detector, any effects near the source no longer alter the path of the photons as they approach the detector.
At which point the trajectories of the photons will be straight/radial from the source and orthogonal to the detector plane when incident (i.e. the rays are parallel to their neighbours). 
Hence, the path of the photons can be reversed by beginning rays at the detector, setting them orthogonal to it, and then propagating them backwards in time to find the location of the photons from the source.
This has the benefit of only considering photons that converge on the detector plane.
Because of this, the observer-to-emitter scheme provides a less computationally heavy workload, as the number of photons traced is significantly smaller than that of the emitter-to-observer scheme 
(where photons begin at the the source and are propagated forwards, counting only those that reach the detector). 
However, the downside of this method restricts the results to a predefined viewing angle of the source for each simulation.
It is easily understood that the viewing angle can alter the power received from a source, especially as the number of symmetries present of the photon source decreases.

Specifically and simply for the work here, photons are sourced at a distant detector plane and back-propagated onto the conversion surface.
During which the path of the photons are effected due to the refraction induced by the plasma and the curvature produced by the NS's mass.
The method we use to complete this scheme is adapted from Refs.~\cite{leroyRadioSignalAxionphoton2020, battyeRadioLineProperties2021, battyeSearchingTimeDependentAxion2023}.
A short list of the process is outlined as follows;
\begin{itemize}
    \item Initialise a 2D detector plane (or image plane) with sufficient separation from the source
    \item Divide the detector plane into pixels of side length $\Delta b$. 
    \item Sourced a single photon at the centre of each pixel.
    \item Propagate each photon using one of the ray tracing methods with a chosen dispersion relations. 
    \item The ray path is then solved by integrating the ray tracing equations over time (or affine parameter) to track the refraction of the rays through the plasma.
    \item Integration is finished when either: (1) the photon intercepts the axion-photon conversion surface; or (2) the photon misses the surface and reaches the end of the integration interval.
\end{itemize}
}

From here, the value of the radiated power received by each pixel can be calculated with the photon values at the conversion surface. The total differential power 
$d{P}/(d\Omega d\omega)$
received by the detector is given by Eqn.~(11) of \citelink{mcdonaldGeneralizedRayTracing2023}{MW23},
\begin{equation}
    \label{eqn:radiated_power}
    \frac{d{P}}{d\Omega d\omega} = \sum_{i,j}\Delta b^2 \omega^3 f^{i,j}_\gamma,
\end{equation}
where the sum over $i, j$ is the index of the pixels on the detector plane, $\Delta b$ is the pixel side length (so that $\Delta b^2$ is the pixel's area), $\omega$ is the frequency/energy of the photon, and $f^{i,j}_\gamma$ is the phase space factor of the photons. 
The photon phase space factor can be related to the axion phase space factor via the axion to photon conversion probability so that $f_\gamma = P_{a\gamma\gamma}f_a$.

We wish to evaluate the quantity \eqref{eqn:radiated_power} at the conversion surface for each photon that intercepts. 
To do this the energy of the photon, the conversion probability and the axion phase space at the location of conversion need to be known. The energy can be found via ray tracing, the conversion probability is found using \eqref{eqn:conversion_probability_isotropic}, and the phase space factor of the axions is given by the expression (e.g. see \citelink{mcdonaldGeneralizedRayTracing2023}{MW23} Eqns.~(58) and (59)),
\begin{equation}
    \label{eqn:axion_phase_space}
    f_a(\vect{x}, \vect{k}) = v_a n_{a,\infty}\frac{k_c(|\vect{x}|)}{k_0}\frac{\delta(\omega-\omega_c)}{4\pi |\vect{k}|^2},
\end{equation}
where $v_a$ is the velocity of the axions at that point, $n_{a,\infty}=\rho_{a,\infty}/m_a$ is the asymptotic number density of axions, $k_c(|\vect{x}|)^2 = k_0^2 + 2GM_{\mathrm{ns}}m_a^2/|\vect{x}|$ is the square of the in-fall momentum of the axions, $k_0 = m_a v_0$ is the momentum dispersion, and $\omega_c$ is the energy of the photon at the conversion surface. 
In natural units, $|\vect{k}|^2=k_\mu k^\mu$ is the three-momentum magnitude of the axion/photon and is found via ray tracing.

\subsubsection{Implementation}

In the following, we describe our implementation of the observer-to-emitter method in GR using the Schwarzschild metric. 
Our code is publicly available on GitHub in Ref.~\cite{satherley_2025_14927160}.
We will use the Schwarzschild coordinate system where $x^i=(t, r, \theta, \phi)$. 
Hence, the 4-momentum will take on the form $k^i=(\omega, k^r, k^\theta, k^\phi)$.
Throughout this description, we will also highlight code variables in our algorithm. 
This is done to clarify what a variable controls and the associated values chosen in this paper. 

The detector's observing angle $(\theta, \phi)$ must be chosen for each simulation. 
This gives the centre line of the detector relative to $\unitvect{z}$ of the NS (recall that $\unitvect{z}$ is in line with the rotation axis of the star). 
These angles are related to the code variables {\bf{Obs\_theta}} and {\bf{Obs\_phi}}. 
The detector plane distance from the NS also needs to be assigned to {\bf{Obs\_r0}}.
A balance must be struck with this initial distance. 
It should be far enough away to approximate what an extremely distant observer would see but not so far that computation time is significantly increased.
The dispersion relationship must also be picked via {\bf{dispersion\_relation}}, where the method is chosen using a string.
Lastly a metric must be chosen using {\bf{metric\_choice}} with the option chose again with a string.

\begin{figure}
    \centering
    \includegraphics[width=1\linewidth]{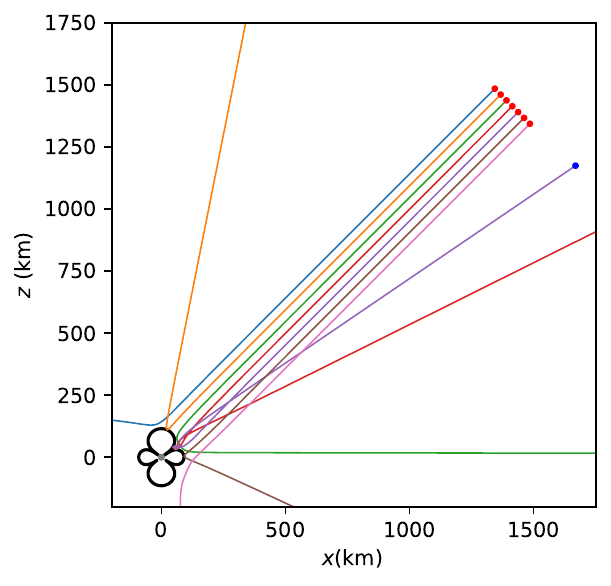}
    \caption{2D representation of the numerical ray tracing in an isotropic plasma. 
    Diffraction due to the plasma causes the deflection of photons.
    The red dots represent the detector plane pixels, where the photons are sourced.
    The coloured lines show the paths of the back-traced photons.
    The blue points denote the end of numerical integration.}
    \label{fig:O2E_ray_paths}
\end{figure}

Below, we outline the procedure used to produce an axion-photon simulation. 
We used the GR dispersion relations in Sec.~\ref{sec:GR_dispersion_relations} and geometric optics in Subsec.~\ref{subsec:GR_ray_tracing}.
A detailed explanation of the observer-to-emitter algorithm used in this article is as follows:
\begin{enumerate}[label=\arabic*)]
    \item The centre of a detector plane is initialised by setting a distance  {\bf{Obs\_r0}} from the NS, and viewing angles {\bf{Obs\_theta}} and  {\bf{Obs\_phi}}. 
    Pixels are then evenly spaced out on a rectangular detector plane, up to a max size of {\bf{max\_x}} and {\bf{max\_y}} with the number of pixels along each row and column given by {\bf{total\_resolution}}.
    The centre of the detector is aligned with the centre of the NS.
    The distance between adjacent pixels is given by $\Delta b =\;${\bf{max\_x}}$/${\bf{total\_resolution}}.

    \item A single photon is then sourced at the centre of each pixel and given initial data:
    \begin{enumerate}
        \item The initial position of each photon is set as the centre of each pixel and given a time {\bf{t\_0}}, the starting time for the simulation (typically $t_0=0$).
        Hence, each photon has an initial 4-position given by,
        \begin{equation}
            x^i=(t_0, r_{\text{pixel}}, \theta_{\text{pixel}}, \phi_{\text{pixel}}),
        \end{equation}
        where pixel refers to the pixel at which the photon is sourced.
        
        \item The initial 4-momentum of the photon is found using the photon energy and the refractive index.
        Energy conservation with the axion gives the energy of each photon during the conversion. 
        All the axion's energy is converted to the photon's energy. 
        One may then use $E^2=m^2+p^2$,
        \begin{equation}
            \label{eqn:O2E_photon_energy}
            \omega=\sqrt{m_a^2\left[1+v_a^2(r_{\text{pixel}})\right]}.
        \end{equation}
        where $\omega$ is the photon's energy, $v_a^2=v^2_{\text{min}}+v_\infty^2$ with $v^2_{\text{min}}=r_s/r$ and $v_\infty$ the dark matter velocity at asymptotic infinity. %
        The equation ~\eqref{eqn:O2E_photon_energy} must be translated via the effective energy equation $\omega=-k_iU^i$ to give the covector form $k_i$. 
        In the case that the plasma is static, such that $U^t=\sqrt{-g^{tt}}$ and $U^\mu=0$, we have that 
        \begin{equation}
            k_t=-\omega/\sqrt{-g^{tt}},
        \end{equation}
        where $\omega$ is given by \eqref{eqn:O2E_photon_energy}.
        
        \item For the spatial components $k_\mu$, we take the refractive index of the medium at distance {\bf{Obs\_r0}} to be $n\simeq1$.
        Hence, we will have the covariant relationship from the refractive index \eqref{eqn:GR_refractive_index} as $-k_tk^t=k_\mu k^\mu$.
        The centre-most photon of the detector is taken to travel perfectly radial from/towards the star at the detector, such that $k^\alpha=(k^r, 0, 0)$.
        So we will have that,
        \begin{equation}
            \label{eqn:O2E_photon_momentum}
            k^r = \sqrt{-g_{tt}k^tk^t/g_{rr}},
        \end{equation} 
        for this photon. %
        Every other photon is assigned the same value to its spatial components but rotated such that the momentum is parallel to the centre-most photon's momentum\footnote{If this rotation is not done, and instead the momentum of each photon is set identical to \eqref{eqn:O2E_photon_momentum}, the photons will all have velocities towards the centre of the star rather than generating rays that are initially parallel.}.
    \end{enumerate}
    
    \item Once the initial conditions of the photons are defined, they can then be back-propagated through the medium using the ray tracing equations \eqref{eqn:GR_geometric_equations_a} and \eqref{eqn:GR_geometric_equations_b} along with one of the dispersion relationships \eqref{eqn:vacuum_dispersion_relation_GR}, \eqref{eqn:unmagnetised_dispersion_relation_GR} or \eqref{eqn:magnetised_dispersion_relation_GR}.
    This is done by using solve\_ivp from Scipy's library.
    An example of the paths that the rays will propagate along during this algorithm is shown in Fig.~\ref{fig:O2E_ray_paths}.
    The figure also shows the origin from which the photons are sourced on the detector plane with red dots, and the end of the integration with blue dots.
    
    \item At the beginning of the simulation, a coarse pixel search over the entire detector plane is done. 
    This search uses the detector resolution\footnote{Resolution here means the number of pixels along the side length of the detector.} defined by {\bf{coarse\_resolution}}.
    The intention of this is to find the regions of the image plane that likely receive photons from the conversion surface.
    This decreases the total number of photons that need to be propagated during the higher resolution search in the next step.
    It is implemented by removing regions of the detector that will not have photons intercepting the conversion surface. %
    Hence, this procedure decreases the run-time and increases the efficiency of the simulation by removing unnecessary ray tracing of photons.
    
    \item After the coarse search has identified fine pixels that may have photons that will back-trace onto the conversion surface, a fine search is started over the smaller pixels. 
    This is done by selecting each coarse pixel identified before and completing a higher-resolution search of that coarse pixel. 
    The resolution is defined by {\bf{fine\_resolution}}.
    Each back-traced photon from a fine pixel that intercepts the conversion surface is recorded, while all photons that never reach the conversion surface are ignored.
    
    \item Numerical integration of a photon's path ceases when either the photon intercepts the NS surface or the end of the integration interval is reached.
    The integration will continue when intercepting the conversion surface
    In the case of these intercepts, the solver records the photon position and momentum 4-vectors so that they can be used to find the power from each intersection of the photon with the conversion surface.

    \item Each photon from the fine search that intercepts the conversion surface can then have its probability of conversion calculated using the information obtained from the ODE solver. 
    Then the photon values and the conversion probability can be used to find the radiated power received by each fine pixel using the terms inside the summation of \eqref{eqn:radiated_power}.
    
\end{enumerate}
Once the algorithm is completed, the results can be plotted to form the image on the detector. 
Each radiated power found in the last step is assigned to the index of the fine pixel from which the photon originated.
These indices can be used to reproduce the image on the detector plane using appropriate plotting software.
Otherwise, the powers for a particular viewing angle can be summed together using \eqref{eqn:radiated_power} to find the total estimated power received by a distant observer.

To improve the effectiveness of the coarse search and avoid missing regions that may receive photons, a relative tolerance {\bf{coarse\_search\_rel\_tol}} is defined in the search algorithm. 
This alters the search so that coarse pixel photons that approach near the conversion surface, but do not intercept it, are also included. 
This is to avoid skipping fine pixels near the edge of a coarse pixel when the centremost\footnote{As the centremost fine pixel produces the same initial photon data as the coarse pixel} fine pixel does not reach the conversion surface.

\section{Results and Discussion} \label{sec:results_and_discussion}

{\renewcommand{\arraystretch}{1.5}
\begin{table}
    \begin{ruledtabular}
        \begin{tabular}{lcc}
        \textrm{Parameter}& \textrm{Symbol}& \textrm{Value}\\
        \colrule
        Axion mass & $m_a$ & $1\microeV$ or $10\microeV$ \\
        Axion-photon coupling constant & $g_{a\gamma\gamma}$ & $\SI{1e-12}{\per\giga\eV}$ \\
        Neutron star mass & $M_{\mathrm{ns}}$ & $M_\odot$ or $2.2M_\odot$ \\
        Neutron star period &$P_{\mathrm{ns}}$ & $2\pi\,\SI{}{s}$ \\
        Neutron star radius &$R_{\mathrm{ns}}$ & $\SI{10}{km}$ \\
        Magnetic field strength at pole &$B_{\mathrm{ns}}$ & $\SI{e10}{T}$ \\
        Misalignment angle & $\chi$ & $0^{\circ}$ \\
        Dark matter dispersion velocity & $v_0$ & $\SI{200}{\km\per\s}$\\
        Dark matter local density & $\rho_{a,\infty}$ & $\SI{0.3}{\giga\eV\per\cm^3}$\\
        Initial simulation distance & $r_{\mathrm{obs}}$ & $200R_{\mathrm{ns}}$ \\
        Coarse-resolution size &%
        & $25$ pixels \\
        Fine-resolution size & %
        & $10$ pixels \\
        \end{tabular}
    \end{ruledtabular}
    \caption{The values for the parameters that define both NS models and our simulation choices. 
    The NS and dark matter parameters match the ones used by \cite{mcdonaldGeneralizedRayTracing2023}.
    However, the final three simulation parameters are chosen by us.
    We use these values in all our simulations unless otherwise stated.
    }
    \label{tab:simulation_parameters}
\end{table}}

We carry out the observer-to-emitter scheme as detailed above using the parameters in Table~\ref{tab:simulation_parameters}.
Because the effects of GR become greater closer to the Schwarzschild radius, we also include a higher mass NS of $M_{\mathrm{ns}}=2.2M_\odot$ (compared to the typical $M_{\mathrm{ns}}=M_\odot$).

We specify the resolution of the detector plane in Table~\ref{tab:simulation_parameters}.
{The number of pixels was chosen by increasing the detector resolution used in trial simulations until the total radiated power had converged. 
This was done for a few chosen observing angles, before a complete run of all observing angles. 
Convergence was checked for the detector angles which receive significant flux from the throat regions of the conversion surface, where higher resolutions are required for resolving the power precisely.}
Hence, our choice to use a total resolution of $250 \times 250$\footnote{The total resolution of the detector in each dimension equals the number of coarse resolution pixels multiplied by the number of fine resolution pixels.} pixels provided a good balance of accuracy and simulation runtime, whilst having converged to a consistent radiated power. 
We found that using fewer pixels than our choice of $250$ did not sufficiently alter the radiated power received in the simulation. 
{Of note, in the higher mass axion and NS simulations, when the conversion surface becomes more compact, a higher detector resolution than $250$ is required.
For our simulation using $m_a=10\microeV$ and $M_{ns}=2.2M_\odot$, a detector resolution of $500 \times 500$ was necessary for the power to converge.}

For the differing axion masses, the conversion surface changes size (e.g. see Fig.~\ref{fig:conversion_surface_2D}). To maintain a resolution of a constant number of pixels, we therefore have to use different pixel sizes for different axion masses.
In the case of $m_a=10\microeV$, the pixel size is $\Delta b=\SI{120}{\metre}$.
For simulations using $m_a=1\microeV$, it has $\Delta b=\SI{480}{\metre}$.

\subsection{GJ vs GLP Conversion Surface}

\begin{figure*}
    \centering
    \includegraphics[width=0.45\textwidth]{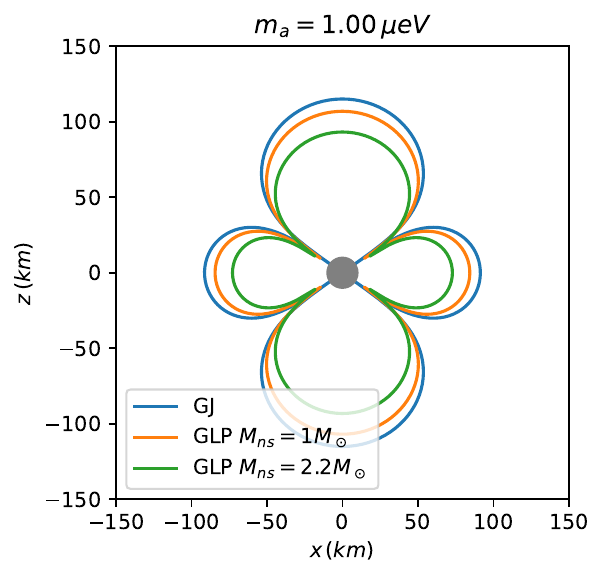}
    \includegraphics[width=0.45\textwidth]{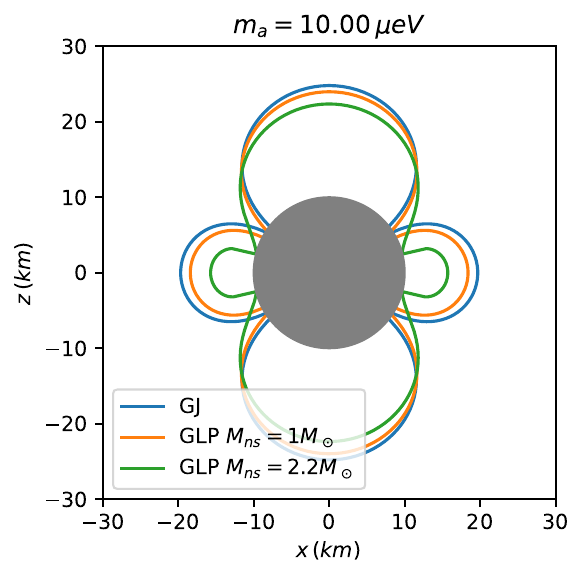}
    \caption{Plots of an xz cross-section (observer at $\theta=90^\circ$) of the surface where axion-photon resonant conversion occurs ($\omega_p=m_a$) showing the two magnetosphere models studied in this paper. 
    For the GLP model, two NS masses are given. 
    The GJ model is unaffected by the star's mass.
    The central grey circle is the NS's surface.
    }
    \label{fig:conversion_surface_2D}
\end{figure*}

Firstly, we explore the effect on the conversion surface.
Each magnetic field choice alters the charge density, hence the plasma frequency and ultimately the point at which the resonance condition $\omega_p=m_a$ is satisfied. 
A 2D cross-section of various conversion surfaces using the GJ and GLP models is shown in Fig.~\ref{fig:conversion_surface_2D}. 
This plot clearly shows that the conversion surface moves closer to the surface of the star when considering the GLP model. 
However, the general regions of the conversion surface remain present, with a bulb at either pole and a central torus around the equator. 
These regions of the conversion surface are separated by throats, which extend into the surface of the NS. 
The throats occur due to a change in sign of the charge density, meaning the plasma frequency tends to zero in the region by the throats.
Hence, from \eqref{eqn:conversion_surface}, the conversion surface radius also tends to zero.

We also display the effect of altering the NS's mass in \figureRef{\ref{fig:conversion_surface_2D}}.
The effects of spacetime curvature are enhanced when the mass of the NS is increased, and hence, there is a greater difference between the two magnetosphere models.
This is because the GLP model depends on the mass of the NS\footnote{The GJ model has no dependence on the mass of the NS as it is derived in flat space. 
However, even in that case, the NS mass will still affect the ray tracing by altering the spacetime geometry.}, ultimately affecting the charge density and, hence, the plasma frequency that gives the conversion surface. 
The effect of increasing the NS mass is even clearer for the higher axion mass as the conversion surface is brought closer to the NS when choosing a heavier axion mass.
This is shown for $m_a=10\microeV$ in the right-hand figure of \figureRef{\ref{fig:conversion_surface_2D}} as opposed to the lighter mass axion used in the left-hand figure.
Most notably, the throats' shape, position, and size are altered more for the higher mass NS and $m_a=10\microeV$.

\subsection{Isotropic Plasma}

\begin{figure}
    \centering
    \includegraphics[width=0.45\textwidth]{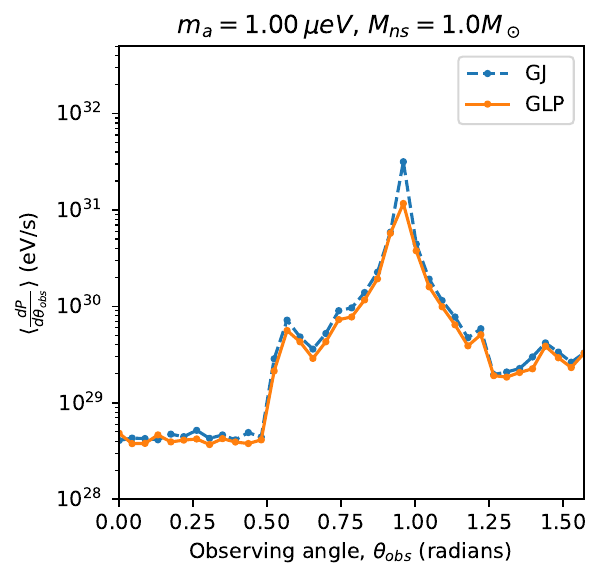}
    \includegraphics[width=0.45\textwidth]{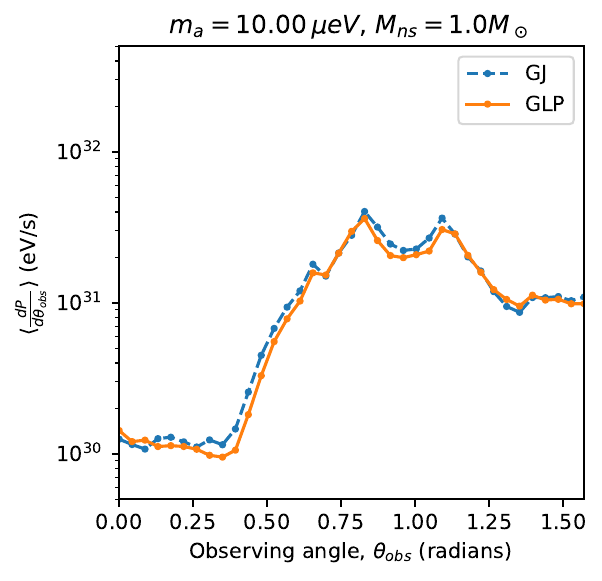}
    \caption{The results of the simulations presented as the period averaged differential power received at an observing polar angle of $\theta_{\text{obs}}$ to the rotation axis of the NS.
    These simulations used an isotropic plasma and the Schwarzschild metric.
    The blue line shows the results from a simulation using the GJ model, whereas the orange lines show the results for the same simulation but instead use the  GLP model.
    }
    \label{fig:O2E_GJ_vs_GR}
\end{figure}

By using the simpler isotropic plasma case \eqref{eqn:unmagnetised_dispersion_relation_GR}, we can glimpse any initial difference in the total power a distant observer receives. 
The main effect of changing the magnetosphere model will be on the charge density, and hence, through \eqref{eqn:conversion_surface}, the plasma frequency around the NS.
Ultimately, this will cause the back-traced photons to have altered trajectories through the plasma.
This will also result in a change to the values returned by \eqref{eqn:conversion_probability_isotropic}, \eqref{eqn:radiated_power} and \eqref{eqn:axion_phase_space} at the point of conversion for a photon.

\begin{figure}
    \centering
    \includegraphics[width=0.45\textwidth]{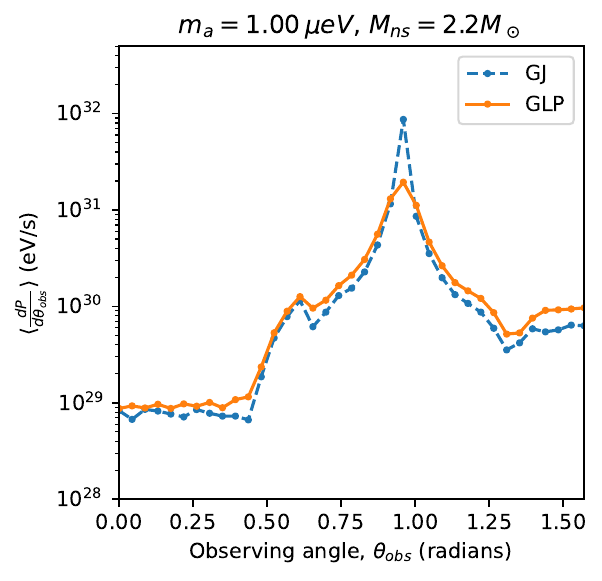}
    \includegraphics[width=0.45\textwidth]{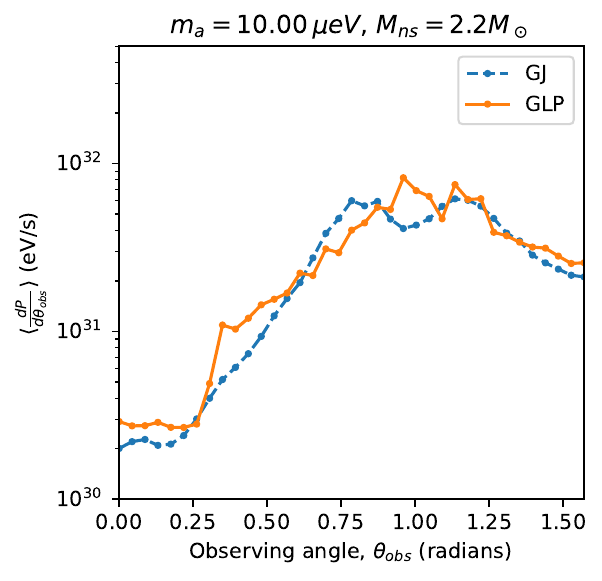}
    \caption{The same as Fig.~\ref{fig:O2E_GJ_vs_GR} but using a NS with a mass $M_{\text{ns}}=2.2M_\odot$.
    For the axion mass $m_a=10\microeV$ simulation using the GLP magnetosphere, the resolution of the detector was set to a higher resolution of $500$ to accurately resolve the total radiated power.
    }
    \label{fig:O2E_GJ_vs_GR_2_2}
\end{figure}

The result of switching the magnetosphere model on the period averaged radiated power across different observing angles is shown in Figs.~\ref{fig:O2E_GJ_vs_GR} and \ref{fig:O2E_GJ_vs_GR_2_2}.
The latter figure uses a higher mass NS in the simulations.
For the results presented in Fig.~\ref{fig:O2E_GJ_vs_GR}, we see only minor differences between the GJ and GLP models.
However, by increasing the mass of the NS, the changes introduced by spacetime curvature will become more important.
Most importantly, increasing the mass significantly alters the GLP magnetosphere model while leaving the GJ magnetosphere model unchanged. 
Changing the mass of the NS also has consequences on the ray tracing due to the changes in the metric. %
Also, the axions will have a greater momentum through the plasma as seen in \eqref{eqn:axion_phase_space}.
The result of all of this can be seen in Fig.~\ref{fig:O2E_GJ_vs_GR_2_2}, that by increasing the mass of the NS, the total radiated power is increased.
It also increases the difference between the two magnetosphere models, especially in the $m_a=10\microeV$ case.
This suggests that a GR magnetosphere may be important to consider in the results of searches around a higher-mass NS. 

\begin{figure}
    \centering
    \includegraphics[width=0.45\textwidth]{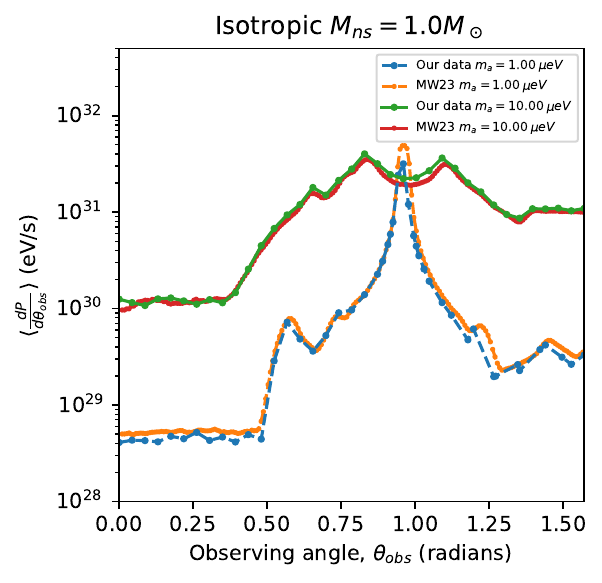}
    \caption{Comparison between our results and those of MW23 for an isotropic plasma using the GJ magnetosphere and Schwarzschild metric. 
    The results from MW23 and our simulations use the parameters in Table~\ref{tab:simulation_parameters}.
    }
    \label{fig:MW23_comparison}
\end{figure}

As a test to ensure that we have implemented the physical processes occurring correctly in our numerical simulations, we compare the results we obtain with the GJ magnetosphere to that of the previously published work of \citelink{mcdonaldGeneralizedRayTracing2023}{MW23}. 
We see in Fig.~\ref{fig:MW23_comparison} that our results reproduce theirs 
reasonably well.
The slight deviation is likely due to differences in numerical solver tolerances and the resolution of the detector plane. 
We also only need to evaluate the polar angles from $0$ to $\pi/2$~radians due to the symmetry of the magnetic field.

To study whether there is a reasonable difference between the total radiated power of the GJ and GLP models, we present the absolute percentage difference between our results using the GJ and GLP models in Figs.~\ref{fig:Power_Difference_Mns_1.0} and \ref{fig:Power_Difference_Mns_2.2}.
To justify if the difference between the two models is significant, we use the difference between our simulations using the GJ magnetosphere and the results from \citelink{mcdonaldGeneralizedRayTracing2023}{MW23}.
This provides an estimate of `uncertainty' in implementing the two models.

In both Figs.~\ref{fig:Power_Difference_Mns_1.0} and \ref{fig:Power_Difference_Mns_2.2}, the simulations using $m_a=1\microeV$ have large differences present at $\theta_{\rm obs}\sim 1$~radian which coincides with the substantial flux received from the throat of the conversion surface.
When considering the conversion surfaces in Fig.~\ref{fig:conversion_surface_2D}, the throats are not as deep for the GLP case. 
Hence, back-traced photons will not `bounce' off the conversion surface as much down its throat, yielding less radiated power.
In the $m_a=10\microeV$ simulations, this issue is not prevalent due to the throat not being as deep and intercepting the NS surface much closer to the opening. 
Interestingly, the interception of the NS surface leads to a slight dip in radiated power near $\theta_{\text{obs}}\sim 1\,\text{rad}$ in the bottom panel Fig.~\ref{fig:O2E_GJ_vs_GR}.

{\renewcommand{\arraystretch}{1.2}
\begin{table}[]
    \centering
    \begin{tabularx}{\columnwidth}{XXXX}
        \hline
        \hline
        Models & $m_a$ & $M_{\text{ns}}$ & Difference \\
        \hline
        \multirow{4}{*}{GLP vs GJ} &  $1\microeV$ & $1.0M_\odot$ & 20\%\\
        &  $1\microeV$ & $2.2M_\odot$ & 32\%\\
        &  $10\microeV$ & $1.0M_\odot$ & 13\%\\
        &  $10\microeV$ & $2.2M_\odot$ & 22\%\\
        \hline
        \multirow{2}{*}{MW23 vs GJ} &  $1\microeV$ & $1.0M_\odot$ & 15\%\\
        &  $10\microeV$ & $1.0M_\odot$ & 9.8\%\\
    \end{tabularx}
    \caption{The average absolute percentage difference between different models with different axion and NS masses across all observing angles.
    The values are found by taking the average of the corresponding lines shown in Figs.~\ref{fig:Power_Difference_Mns_1.0} and \ref{fig:Power_Difference_Mns_2.2}.
    }
    \label{tab:percentage_difference}
\end{table}

In Fig.~\ref{fig:Power_Difference_Mns_2.2}, for the $m_a=1\microeV$ and $m_a=10\microeV$ cases, respectively, a reasonable difference is present between the GJ and GLP models across most viewing angles.
This difference is also greater than the difference between GJ and \citelink{mcdonaldGeneralizedRayTracing2023}{MW23} data.
For the GLP model, from Table~\ref{tab:percentage_difference}, there is an average absolute difference of $32\%$ and $22\%$ over all the viewing angles, which, when compared to the difference with \citelink{mcdonaldGeneralizedRayTracing2023}{MW23} having an average of $15\%$ and $9.8\%$, appears to be a significant change in power.
Looking closer at Fig.~\ref{fig:Power_Difference_Mns_2.2}, we see that some viewing angles have larger changes in power, while a few have a minimal difference.

In Fig.~\ref{fig:Power_Difference_Mns_1.0} for both axion masses, the difference between the GJ and GLP models is similar to the difference between GJ and \citelink{mcdonaldGeneralizedRayTracing2023}{MW23}.
Hence, the GLP model does not introduce a significant difference in the lower mass NS case.
Hence, for a higher mass NS and a conversion surface that will be close to the Schwarzschild radius, a GR magnetosphere induces a reasonable difference in power.

\begin{figure}
    \centering
    \includegraphics[width=0.45\textwidth]{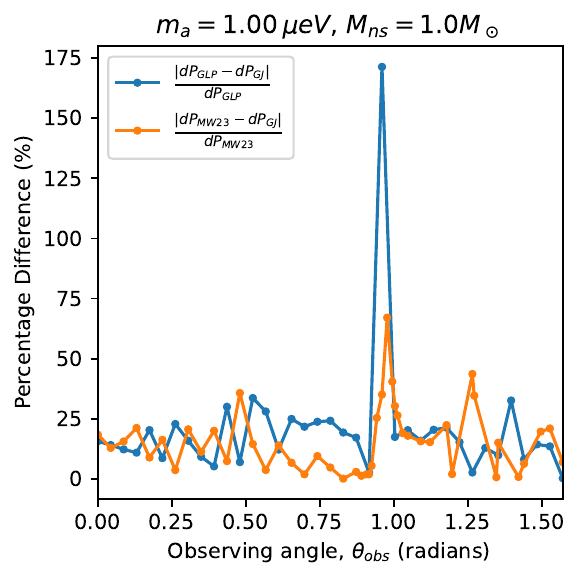}
    \includegraphics[width=0.45\textwidth]{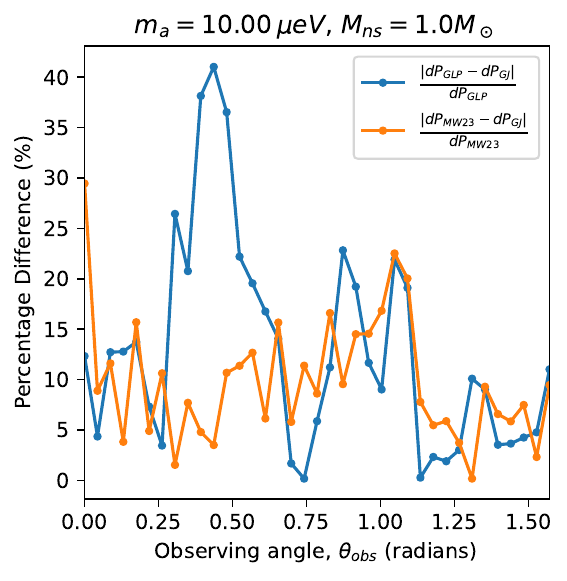}
    \caption{The difference in averaged power between different numerical simulations.
    The blue line is the difference between our simulations using the GJ and GLP models.
    The orange line is the difference between our simulation and the simulation data from MW23.
    We use the orange line as a simple estimate of the uncertainty in our blue line.
    }
    \label{fig:Power_Difference_Mns_1.0}
\end{figure}

\begin{figure}
    \centering
    \includegraphics[width=0.45\textwidth]{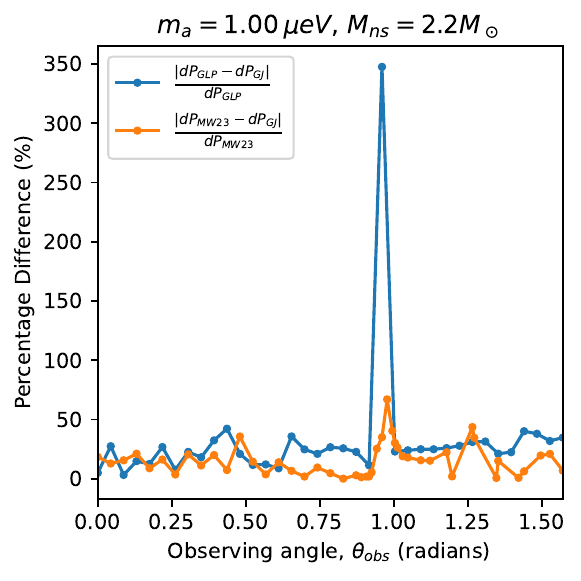}
    \includegraphics[width=0.45\textwidth]{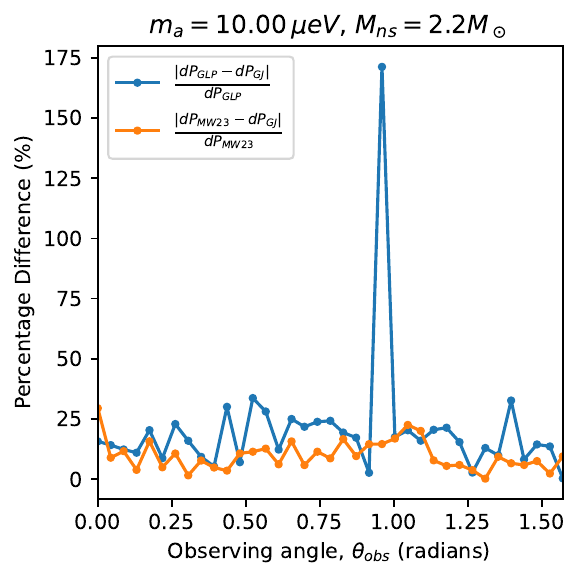}
    \caption{Same as Fig.~\ref{fig:Power_Difference_Mns_1.0} but for a NS mass of $M_{\mathrm{ns}} = 2.2M_\odot$. 
    Note that the simulation data from MW23 is still for a NS mass of $M_{\mathrm{ns}} = M_\odot$.
    }
    \label{fig:Power_Difference_Mns_2.2}
\end{figure}


\section{Conclusion}

In the work carried out here, we discussed the implementation of the GLP magnetosphere model that is derived in a curved spacetime using the Schwarzschild metric. 
This GR model was then employed in numerical simulations to study the effect on axion-photon conversion signals from NS.
This is in comparison to the recent numerical simulations of \citelink{mcdonaldGeneralizedRayTracing2023}{MW23}, which uses the flatspace-derived GJ model in their simulations.
Using an isotropic dispersion relation, we compared the difference in total radiated power across the observing angles of the neutron star between the two models that were simulated for this article and the simulation data supplied by \citelink{mcdonaldGeneralizedRayTracing2023}{MW23}.

For the $M_{\rm ns}=2.2M_\odot$ case, we found the average absolute percentage difference between the GLP and GJ models was 22\% for the $m_a=10\mu$eV. This was about 2.2 times greater than the average difference between our GJ model and that of \citelink{mcdonaldGeneralizedRayTracing2023}{MW23}. The $m_a=1\mu$eV had an even larger 32\% difference between the GLP and GJ models, which was 2.1 times greater than the average difference between our GJ model and that of \citelink{mcdonaldGeneralizedRayTracing2023}{MW23} for this case.

In the cases of $M_{\rm ns}=M_\odot$, implementing the GLP model appears to have little effect on the radiated power for an isotropic plasma.
The conclusion that the higher mass case has the greatest difference intuitively makes sense.
This is due to the stronger dependence on spacetime curvature effects when the conversion surface is close to the Schwarzschild radius of the NS.

Earlier in this article, we discussed the anisotropic plasma relationships \eqref{eqn:magnetised_dispersion_relation_GR}. 
However, our implementation of this more complex plasma (see Ref.~\cite{gedalinWavesStronglyMagnetized2001}) did not produce results that reliably matched the results presented by \citelink{mcdonaldGeneralizedRayTracing2023}{MW23} for the GJ model. 
Hence, we have excluded any of our results using the dispersion relation \eqref{eqn:magnetised_dispersion_relation_GR}.
In future work, we will determine the reason for this difference in results. 
{Most importantly, this had no effect on our reproduction of their isotropic plasma results.
The exclusion of an anisotropic plasma does not alter the conclusion in this foundational model, that in this simpler scenario, the GLP model causes a difference in the total power received.

However, by including an anisotropic plasma, the ray tracing of photons through the plasma will have a greater dependence on the magnetic field model.
Hence, with this in mind, we suspect that the difference in power between the GJ and GLP models would increase further when considering an anisotropic plasma.
The results of this paper provide a necessary computational step toward understanding whether such extended simulations will yield further insights, which we leave for future work.

Other elements which will further alter the radiated power include using an inclined magnetic field and adding more multi-pole components beyond the dipole (see Ref.~\cite{grallaInclinedPulsarMagnetospheres2017a}).
Both of which will alter the magnetosphere surrounding the NS, changing the ray paths, conversion surface, and the conversion probabilities.
All of which has the potential to change the power found in simulations of axion to photon conversion.
Most notably, by inclining the magnetic field, the radiated power will become dependent on the phase of the rotation (and the azimuthal angle).
Meaning a simulation would need to complete a full rotation of the NS to obtain the period-averaged power.
This is as opposed to an aligned rotator, where the power is constant throughout the entire rotation due to the symmetry present.

Lastly, in the situation studied here, we provide an initial step at including a magnetosphere model that incorporates GR.
However, this entertains the idea of considering more complex NS magnetosphere models.
Of particular interest is the use of a numerical General Relativistic Magnetohydrodynamics (GRMHD) simulation to model the NS magnetosphere (such as \href{https://ui.adsabs.harvard.edu/abs/2019ascl.soft12014T/abstract}{HARMPI}).
This different model could produce further changes to the total radiated power in axion to photon conversion simulations.
However, to know if the analytical GLP model we choose here yields sufficient precision, compared to a GRMHD model, requires further investigation.
}

\begin{acknowledgments}
We thank Jamie McDonald, Harrison Ploeg, and Sam Witte for helpful comments. We also thank Sam Witte for making available some of the simulation data from \citelink{mcdonaldGeneralizedRayTracing2023}{MW23} to include in our Figs.~\ref{fig:MW23_comparison}, \ref{fig:Power_Difference_Mns_1.0}, and \ref{fig:Power_Difference_Mns_2.2}.
We also thank Alexandru Lupsasca for helpful comments on the difference between the GLP model's choice of coordinates and our definition of the same coordinates.
\end{acknowledgments}

\section*{Data Availability}
The data that support the findings of this article are openly available \cite{satherley_2025_14927160}.

\bibliographystyle{apsrev4-2-author-truncate.bst}
\bibliography{references} %

\appendix

\section{GLP - Inclined Rotator} \label{appendix:GLP_inclined}

The third paper in the series by Gralla et al.~\cite{grallaInclinedPulsarMagnetospheres2017a}, extends their work by including a misalignment between the rotation and magnetic field axis. 
They suggest that the results from their first paper can be modified by a spatial coordinate change, using a set of spatial coordinates about the rotation axis and another set about the magnetic field axis. 

The rotation axis $\vect{\Omega}$ is chosen to be in line with the Cartesian axis $z$, after which the typical spherical polar coordinates $(r, \theta, \phi)$ are defined about this axis.
This is the coordinate system a stationary observer will be using. 
The magnetic field symmetry axis $\vect{e}$ is inclined by a polar angle $\chi$ to the rotation axis and will have azimuthal angle $\Omega t$ due to the star's rotation. 
The polar coordinates $(r, \vartheta, \varphi)$ are defined around the axis $\vect{e}$ such that $\vartheta$ is the polar angle measured away from the axis and $\varphi$ is the azimuthal angle around the axis measured from a line pointing at the rotation axis $\vect{\Omega}$, such that in this coordinate system the magnetic field is independent of time.
We can also introduce an azimuthal angle, which measures the angle from $\vect{e}$ to $\phi$ in the $(r,\theta,\phi)$ frame. Explicitly, this can be expressed as $\lambda=\phi-\Omega t$.

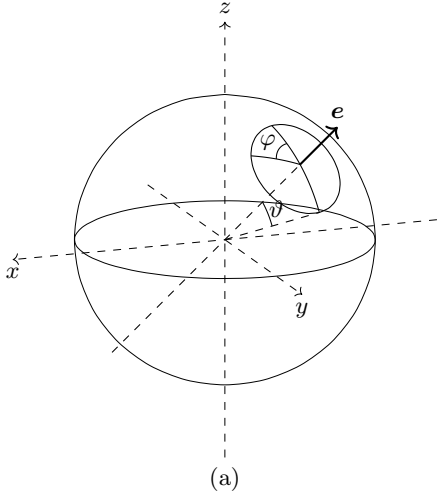
\begin{figure}
    \centering
    \begin{tikzpicture}
        \begin{scope}[viewport={160}{15}, very thin]
        
            \draw[dashed,->] (\ToXYZr{-1.5}{90}{0}) -- (\ToXYZr{1.5}{90}{0}) node[anchor=north]{$x$};
            \draw[dashed,->] (\ToXYZr{-1.5}{90}{90}) -- (\ToXYZr{1.5}{90}{90}) node[anchor=north]{$y$};
            \draw[dashed,->] (\ToXYZr{-1.5}{0}{0}) -- (\ToXYZr{1.5}{0}{0}) node[anchor=south]{$z$};
            \node at (0, 0, -1.65) {(a)};
            
            \draw[domain=0-\Rotation:360-\Rotation, variable=\azimuth, smooth] plot (\ToXYZ{90}{\azimuth});
            \draw[domain=0:360, variable=\elevation, smooth] plot (\ToXYZ{\elevation}{\Rotation});
        \end{scope}
        \tdplotsetmaincoords{0}{0}
        \tdplotsetrotatedcoords{45}{45}{-112.5}
        \begin{scope}[tdplot_rotated_coords]
            \draw[domain=-3:2, variable=\radius, dashed] plot (\ToXYZr{\radius}{0}{0});
            \draw[domain=2:3, variable=\radius, thick, ->] plot (\ToXYZr{\radius}{0}{0}) node[anchor=south]{$\vect{e}$};
            
            \draw[domain=0:360, variable=\azimuth, smooth] plot (\ToXYZr{2}{20}{\azimuth});
            \draw[dashed] (\ToXYZr{0}{20}{0}) -- (\ToXYZr{2}{20}{0});
            \draw[domain=0:20, variable=\polar, smooth] plot (\ToXYZr{1}{\polar}{0}) node[] at (\ToXYZr{1.2}{10}{0}) {$\vartheta$};
            \draw[domain=-20:20, variable=\polar, smooth] plot (\ToXYZr{2}{\polar}{0});
            \draw[domain=180:240, variable=\azimuth, smooth] plot (\ToXYZr{2}{10}{\azimuth}) node[] at (\ToXYZr{2}{15}{210}) {$\varphi$};
            \draw[domain=0:20, variable=\polar, smooth] plot (\ToXYZr{2}{\polar}{240});
        \end{scope}
    \end{tikzpicture}
    \caption{The relationship between $(r,\vartheta,\varphi)$ and the Cartesian coordinates $x$, $y$, and $z$. 
    This shows the coordinate system $(r,\vartheta,\varphi)$ portrayed on a sphere.
    The angle $\vartheta$ is measured as a polar angle from $\vect{e}$, and the angle $\varphi$ is measured anticlockwise from a line pointing towards $z$.}
    \label{fig:vartheta_varphi}
\end{figure}

Given axisymmetric field functions defined about the magnetic field symmetry axis, the coordinates $(r, \vartheta, \varphi)$ can be used to maintain the field %
functions' %
symmetry.
These coordinates are shown in Fig.~\ref{fig:vartheta_varphi}. 
Hence, we may take the two functions from Gralla et al. \cite{grallaPulsarMagnetospheresFlat2016} that describe an aligned dipole, \eqref{eqn:mag_flux_function_1} and \eqref{eqn:mag_flux_function_2}, and perform the coordinate change $\theta\rightarrow\vartheta$ and $\phi\rightarrow\varphi$. This results in,
\begin{subequations}
    \begin{align}
            \psi_{1\;\text{near}}(r, \vartheta) &= \mu R^>_1(r)\sin^2\vartheta,\label{eqn:inclined_mag_flux_function_1}\\
            \psi_2(\varphi) &= \varphi \label{eqn:inclined_mag_flux_function_2},
    \end{align}
\end{subequations}
where the time dependence no longer exists in the rotated frame due to the axisymmetric field functions.

The coordinate angles $\vartheta$ and $\varphi$ need to be related to the angles $\theta$ and $\phi$ so that we may compute the fields and charge density in an observer's frame. A method to derive these relations is using spherical triangles, and in particular, the relationships are,
\begin{subequations}
    \begin{align}
        \cos\vartheta &= \cos\theta\cos\chi + \sin\theta\cos\lambda\sin\chi, \label{eqn:cos_vartheta}\\
        \tan\varphi &= \frac{\sin\theta\sin\lambda}{\cos\theta\sin\chi + \sin\theta\cos\lambda\cos\chi}, \label{eqn:tan_varphi}
    \end{align}
\end{subequations}
which are found using the equations and figures in Appendix~\ref{Appendix:SphericalTrangle}.\footnote{These relationships differ from Ref.~\cite{grallaInclinedPulsarMagnetospheres2017a}. 
This is due to different coordinate definitions. 
However, the results remain unaffected.} 
The relationships \eqref{eqn:inclined_mag_flux_function_1}, \eqref{eqn:inclined_mag_flux_function_2}, \eqref{eqn:cos_vartheta} and \eqref{eqn:tan_varphi} can all be combined with \eqref{eqn:electromagnetic_tensor_index_form} to give the electromagnetic tensor around an inclined star with GR corrections due to the curved spacetime.
We can then use this electromagnetic tensor with \eqref{eqn:covariant_magnetic_field} to find the magnetic field strength throughout the star's magnetosphere.

Lastly, we need the charge density for this NS model to provide the plasma. From Ref.~\cite{grallaInclinedPulsarMagnetospheres2017a}, in the $(r, \theta, \phi)$ frame, they give charge density as,
\begin{widetext}
\begin{equation}
    \label{eqn:GR_charge_density_gralla}
    \begin{aligned}
        J^{\hat{t}}=\frac{\Omega-\Omega_z}{r(r-2 G M)}\biggl\{&\partial_\theta \psi_1 \partial_\theta \partial_\phi \psi_2 - \partial_\theta \psi_2 \partial_\theta \partial_\phi \psi_1 + r(r-2 G M)\left(\partial_r \psi_1 \partial_r \partial_\phi \psi_2-\partial_r \psi_2 \partial_r \partial_\phi \psi_1\right) \\
        &-\partial_\phi \psi_1\left[\left(1-\frac{2 G M}{r}\right) \partial_r\left(r^2 \partial_r \psi_2\right)+\frac{\partial_\theta\left(\sin \theta \partial_\theta \psi_2\right)}{\sin \theta}\right] \\
        &+ \partial_\phi \psi_2\left[\left(1-\frac{2 G M}{r}\right) \partial_r\left(r^2 \partial_r \psi_1\right)+\frac{\partial_\theta\left(\sin \theta \partial_\theta \psi_1\right)}{\sin \theta}\right]\biggl\}.
    \end{aligned}
\end{equation}
\end{widetext}
Where \eqref{eqn:GR_charge_density_gralla} was found via \eqref{eqn:4-current_em_tensor} using \eqref{eqn:electromagnetic_tensor_index_form} with the coordinate dependencies induced by \eqref{eqn:cos_vartheta} and \eqref{eqn:tan_varphi}. For a stationary Schwarzschild observer,
the charge density $\rho_e$ is given by,
\begin{equation}
    \rho_e=U_aJ^a=J^t\sqrt{1-\frac{r_s}{r}}=J^{\hat{t}}.
\end{equation}

This gives us a well-defined near-field approximation of a misaligned dipole magnetic field in covariant form which accounts for the curvature of spacetime.

\section{Solutions of Spherical Triangles}
\label{Appendix:SphericalTrangle}
\begin{figure}
    \subfloat[]{
        \centering
        \begin{tikzpicture}
            \begin{scope}[viewport={\Rotation}{\Tilt}, very thin]
                \draw[domain=0:58.5, variable=\polar, smooth] plot (\ToXYZr{2}{\polar}{\Rotation-136.5}) node[] at (\ToXYZr{2.2}{35}{\Rotation-135}) {$a$};
                
                \draw[domain=0:50, variable=\polar, smooth] plot (\ToXYZr{2}{\polar}{\Rotation-70}) node[] at (\ToXYZr{2.2}{30}{\Rotation-65}) {$b$};
                
                \node at (\ToXYZr{2}{41}{\Rotation-85}) {$A$};
                \node at (\ToXYZr{2}{45}{\Rotation-125}) {$B$};
                \node at (\ToXYZr{2}{15}{\Rotation-98}) {$C$};
            \end{scope}
            
            \tdplotsetmaincoords{0}{0}
            \tdplotsetrotatedcoords{-45+\Tilt}{-45}{0}
            \begin{scope}[tdplot_rotated_coords, very thin]
                \draw[domain=0:53, variable=\polar, smooth] plot (\ToXYZr{4}{\polar}{40}) node[] at (\ToXYZr{4}{30}{30}) {$c$};
            \end{scope}
            
        \end{tikzpicture}               
        \label{fig:spherical_triangle}
    }
    \subfloat[]{
        \centering
        \begin{tikzpicture}
            \begin{scope}[viewport={\Rotation}{\Tilt}, very thin]
            
                \draw[dashed,->] (0,0,1) -- (0,0,\VectorLen) node[anchor=south]{$z, \vect{\Omega}$};
                
                \draw[domain=-20:160, variable=\azimuth, smooth] plot (\ToXYZ{90}{\azimuth});
                \draw[domain=160:340, variable=\azimuth, dashed] plot (\ToXYZ{90}{\azimuth});
                \draw[domain=0:360, variable=\elevation, smooth] plot (\ToXYZ{\elevation}{\Rotation});
                
                \draw[domain=0:58.5, variable=\polar, smooth] plot (\ToXYZr{1}{\polar}{\Rotation-136.5}) node[] at (\ToXYZr{1.1}{35}{\Rotation-135}) {$\chi$};
                
                \draw[domain=0:50, variable=\polar, smooth] plot (\ToXYZr{1}{\polar}{\Rotation-70}) node[] at (\ToXYZr{1.1}{30}{\Rotation-65}) {$\theta$};
                
                \node at (\ToXYZr{1}{45}{\Rotation-125}) {$-\varphi$};
                \node at (\ToXYZr{1}{41}{\Rotation-85}) {$\xi$};
                \node at (\ToXYZr{1}{15}{\Rotation-98}) {$\lambda$};

            \end{scope}
            \tdplotsetmaincoords{0}{0}
            \tdplotsetrotatedcoords{-45+\Tilt}{-45}{0}
            
            \def\EVectorFieldPolar{40}
            \def\EVectorFieldAzimuth{225}
            
            \begin{scope}[tdplot_rotated_coords, very thin]
                \draw[domain=2:3, variable=\radius, thick, ->] plot (\ToXYZr{\radius}{0}{0}) node[anchor=south]{$\vect{e}$};
                
                \draw[domain=0:53, variable=\polar, smooth] plot (\ToXYZr{2}{\polar}{40}) node[] at (\ToXYZr{2}{30}{30}) {$\vartheta$};
            \end{scope}
        \end{tikzpicture}
        \label{fig:spherical_triangle_coordinates}
    }
    \caption{Spherical triangles with general notation and the labels according to parameters defined in this article.
    (a) A spherical triangle with side lengths and angles labelled according to the notation in this appendix. 
    The lower case depicts a length, while the upper case depicts an angle.
    Note that sides and angles labelled with the same letter are opposite each other. 
    (b) Description according to coordinate systems $(r, \theta, \phi)$ and $(r, \vartheta, \varphi)$ including the misalignment angle $\chi$ of $\vect{e}$ axis and $\lambda=\phi-\Omega t$ due to the rotation. 
    We take all coordinate angles to increase in the anticlockwise direction about their given axis.
    This is why we take $\varphi$ to increase in the anticlockwise direction looking down $\vect{e}$.
    Hence, in this figure, it has a negative value. 
    The angle $\xi$ is an angle not required in this article.}
\end{figure}
The spherical Law of Cosines is given as (e.g.~\cite{vanbrummelenHeavenlyMathematicsForgotten2013}),
\begin{equation}
\label{eqn:sph_tri_2_sides}
    \cos c = \cos a \cos b + \sin a \sin b \cos C,
\end{equation}
and the spherical Law of Sines is given as,
\begin{equation}
    \label{eqn:sphi_tri_sin}
    \frac{\sin a}{\sin A} = \frac{\sin b}{\sin B},
\end{equation}
where the lower case represents the side length and the upper case represents the corresponding angle (see Fig.~\ref{fig:spherical_triangle}).
Upon rearranging the spherical Law of Cosines for $\cos C$, we get,
\begin{equation}
\label{eqn:sph_tri_3_sides}
    \cos C=\frac{\cos c - \cos a\cos b}{\sin a \sin b}.
\end{equation}
We can replace $a$, $b$, and $c$ with the coordinate angles around the rotation axis $(\theta, \phi)$ and around the magnetic field axis $(\vartheta, \varphi)$ (see Fig.~\ref{fig:spherical_triangle_coordinates}),
\begin{subequations}
    \begin{gather}
        \cos \vartheta = \cos \chi \cos \theta + \sin \chi \sin \theta \cos \lambda,\label{eqn:D.4a}\\
        \frac{\sin\vartheta}{\sin\lambda} = \frac{\sin\theta}{\sin{-\varphi}},\label{eqn:D.4b}\\
        \cos{-\varphi} = \frac{\cos\theta-\cos\chi\cos\vartheta}{\sin\chi\sin\vartheta}.\label{eqn:D.4c}
    \end{gather}
\end{subequations}
All three equations can be combined to remove the dependency on $\vartheta$ from \eqref{eqn:D.4c}. By replacing $\sin\vartheta$ using \eqref{eqn:D.4b} and $\cos\vartheta$ using \eqref{eqn:D.4a}
\begin{equation}
    \cos{\varphi} = \frac{-\sin\varphi[\cos\theta - \cos\chi(\cos \chi \cos \theta + \sin \chi \sin \theta \cos \lambda)]}{\sin\chi\sin\lambda\sin\theta}
\end{equation}
where we have also simplified the negative arguments. This then becomes,
\begin{subequations}
    \begin{align}
        \frac{\cos{\varphi}}{\sin\varphi} &= \frac{-\cos\theta + \cos^2\chi\cos\theta - \cos\chi\sin\chi\sin\theta\cos\lambda}{\sin\chi\sin\lambda\sin\theta},\\
        \tan\varphi &= \frac{\sin\chi\sin\lambda\sin\theta}{\sin^2\chi\cos\theta - \cos\chi\sin\chi\sin\theta\cos\lambda},\\
        \tan\varphi &= \frac{\sin\lambda\sin\theta}{\sin\chi\cos\theta - \cos\chi\sin\theta\cos\lambda}.
    \end{align}
\end{subequations}

\end{document}